\documentclass[12pt]{article}
\catcode`\@=11
\def\numberbysection{\@addtoreset{equation}{section}
\def\theequation{\thesection.\arabic{equation}}}
\numberbysection

\newcommand{\beq}{\begin{equation}}
\newcommand{\eeq}{\end{equation}}
\newcommand{\bea}{\begin{eqnarray*}}
\newcommand{\eea}{\end{eqnarray*}}
\newcommand{\beqa}{\begin{eqnarray}}
\newcommand{\eeqa}{\end{eqnarray}}
\newcommand{\comport}[2]{\mathrel{\mathop{#1}\limits_{#2}^{}}}

\newcommand{\e}{{\rm e}}
\newcommand{\eq}{{\rm eq}}

\renewcommand{\i}{{\rm i}}
\renewcommand{\Im}{\mathop{\rm Im}}
\newcommand{\lap}[1]{\mathrel{\mathop{\cal L}\limits_{#1}^{}}}

\newcommand{\sign}{\mathop{\rm sign}}

\begin{document}
\centerline{\Large\bf Statistics of the occupation time of renewal processes}
\vspace{1cm}
\centerline{\large
by C.~Godr\`eche$^{a,}$\footnote{godreche@spec.saclay.cea.fr}
and J.M.~Luck$^{b,}$\footnote{luck@spht.saclay.cea.fr}}
\vspace{1cm}
\centerline{$^a$Service de Physique de l'\'Etat Condens\'e,
CEA Saclay, 91191 Gif-sur-Yvette cedex, France}
\vspace{.1cm}
\centerline{$^b$Service de Physique Th\'eorique,
CEA Saclay, 91191 Gif-sur-Yvette cedex, France}
\vspace{1cm}
\begin{abstract}
We present a systematic study of the statistics of the occupation time and
related random variables for stochastic  processes with independent
intervals of time. According to the nature of the distribution of time
intervals, the probability density functions of these random variables have
very different scalings in time. We analyze successively the cases where
this distribution is narrow, where it is broad with index $\theta <1$, and
finally where it is broad with index $1<\theta <2$. The methods introduced
in this work provide a basis for the investigation of the statistics of the
occupation time of more complex stochastic processes (see joint paper by G.
De Smedt, C. Godr\`{e}che, and J.M. Luck \cite{ustocome}).
\end{abstract}
\vfill
\noindent JSP 100-147 -- revised
\hfill S/00/037
\vskip -1pt
\noindent P.A.C.S.: 02.50.Ey, 02.50.Ga, 05.40.+j
\hfill T/00/086
\newpage

\setcounter{footnote}{0}
\section{Introduction}

The question of determining the distribution of the occupation time of
simple stochastic processes has been well debated by probabilists in the
past \cite{levy,erdos,kac,darling,lamperti}. 
The simplest example is that of the
binomial random walk, or, in its continuum version, of Brownian motion. 
Let $x_{t}$ be the position of the walker at time $t$, where $t$ is an integer,
for the random walker, or a continuous variable, for Brownian motion. 
Then define the stochastic process $\sigma_{t}=\sign x_{t}$. The
occupation times $T_{t}^{+}$ and $T_{t}^{-}$ are the lengths of time spent
by the walker, respectively on the right side, or on the left side of the
origin, up to time $t$ (using continuous time): 
\begin{equation}
T_{t}^{\pm }=\int_{0}^{t}dt'\frac{1\pm \sigma_{t'}}{2}.
\label{defTt}
\end{equation}
It is also convenient to define the more symmetrical quantity 
\begin{equation}
S_t=tM_{t}=\int_{0}^{t}dt'\sigma_{t'}
=T_{t}^{+}-T_{t}^{-}.
\label{def_S}
\end{equation}
$M_t$, the mean of the stochastic process $\sigma_t$, will
be hereafter referred to as the mean magnetization, 
by analogy with physical situations where $\sigma_t$ is
the spin at a given point of space.

A classical result, due to P. L\'{e}vy, is that the limiting distribution of
the fraction of time spent by the walker on one side of the origin up to
time $t$, as $t\rightarrow \infty $, is given by the arcsine law\footnote{%
See appendix~A for the notation.} 
\begin{equation}
\lim_{t\rightarrow \infty }f_{t^{-1}T^{\pm }}(x)=\frac{1}{\pi \sqrt{x(1-x)}}%
\qquad (0< x< 1)  \label{arcsin1}
\end{equation}
or equivalently, for the mean magnetization, 
\begin{equation}
f_{M}(x)=\frac{1}{\pi \sqrt{1-x^{2}}}\qquad (-1< x< 1).
\label{arcsin2}
\end{equation}

The interpretation of this result is that, contrarily to intuition, the
random walker spends most of its time on one side of the origin. Translated
into the language of a game of chance, this amounts to saying that one of
the players is almost always winning, or almost always loosing, a situation
which can be summarized by the ``persistence of luck, or of bad luck''.

On the other hand, a well-known result is that the probability for the
random walker to remain on one side of the origin up to time $t$, i.e., for
the stochastic process $\sigma_{t}$ not to change sign up to time $t$,
decays as $t^{-\frac{1}{2}}$, which defines the first-passage
exponent~$\frac{1}{2}$. 
The identity between this exponent and that characterizing the
singularity of the probability density functions 
(\ref{arcsin1}, \ref{arcsin2}) for $x\rightarrow 1$ is not coincidental. 
Both exponents are actually
intimately related, as will be explained later in more generality.

The stochastic process $\sigma_t$ defined above is simple in two
respects. Firstly, the steps of the walker, or the increments of Brownian
motion, are independent. Secondly, the process is zero-dimensional, in the
sense that it does not interact with other processes.

The question of {\it phase persistence} for self-similar growing systems,
which appeared more recently in statistical physics, is a natural
generalization of the probabilistic problem presented above. The physical
systems in consideration are for instance breath figures, the pattern formed
by growing and coalescing water droplets on a plane \cite{beysens,breath}; 
systems of spins quenched from high temperature to zero
temperature, or more generally to the low-temperature phase \cite{ising}; or
the diffusion field evolving in time from a random initial condition \cite
{diffusion}. The question posed is: what fraction of space remained in the
same phase up to time $t$, or equivalently, what is the probability for a
given point of space to remain in this phase. In this context, the
stochastic process $\sigma_{t}$ is, for instance, the indicator of whether
a given point of space is in the dry phase for the breath figure experiment,
the spin at a given site of a lattice for the zero-temperature kinetic Ising
model, or the sign of the field at a given point in space for the diffusion
equation.

In these  cases, the probability for a given point of space to remain
in a given phase is equal to the probability $p_{0}(t)$ that the process $%
\sigma_{t}$ did not change sign up to time $t$, or persistence probability.
For these situations, and similar ones, $p_{0}(t)$ decays algebraically with
an exponent $\theta $, the persistence exponent, which plays the role in the
present context of the first-passage exponent $\frac{1}{2}$ of the random
walk.

Phase persistence is far more complex than the simple case of the
persistence of luck for the random walker, because the physical systems
considered are spatially extended. Thus, the stochastic process $\sigma_{t}$%
, which is defined at a given point of space, interacts with similar
processes at other points of space. As a consequence, the persistence
exponent $\theta $ is usually hard to determine analytically.

Pursuing the parallel with the random walk, one may wonder, for the physical
systems where phase persistence is observed, what is the behavior of the
distribution of the occupation time (\ref{defTt}), or of the mean
magnetization (\ref{def_S}). 

For the zero-temperature Glauber-Ising chain \cite{d-g}, the diffusion
equation \cite{d-g,newman}, the two-dimensional Ising model \cite
{drouffe}, or for growing surfaces \cite{newman2}, non-trivial U-shaped
distributions $f_{M}$ are observed. The singularity exponent of these
distributions, as $M_{t}\rightarrow \pm 1$, gives the persistence exponent,
thus providing a stationary definition of persistence, even at finite
temperature \cite{drouffe}. The only analytical results at our disposal for
these examples is for the diffusion equation, using the so-called
independent-interval approximation \cite{d-g}. However, understanding the
origin of the existence of a limiting distribution is easy. In the long-time
regime the two-time autocorrelation is a function of the ratio of the two
times. Therefore the variance of $M_{t}$ remains finite 
(see equation~(\ref{eq_var}) below and ref. \cite{d-g}).

Another case of interest is the voter model. The investigation of the
distribution of the occupation time for this model was introduced by Cox and
Griffeath \cite{cox-gr} (see \cite{liggett} for a summary). 
In contrast
with the former examples, persistence is not algebraic for this case
\cite{ben,howard}, and
the distribution $f_{M}$ is slowly peaking with time
($\langle M_t^2\rangle\sim1/\ln t$).
This is
akin to what occurs in the Ising model quenched from high temperature to the
critical point. 
The explanation is the same in both cases: the two-time
autocorrelation is the product of a scaling function of the ratio of the two
times by a prefactor depending on one of the two times.
In the case of the Ising model, this prefactor
is related to the anomalous dimension of the field at criticality
(see equation (\ref{ctt_ising}) below and ref. \cite{g-l}). 
As a consequence, the variance of $M_{t}$ is slowly decreasing in time
($\langle M_t^2\rangle\sim t^{-2\beta/\nu z_c}$).
Let us
recall that the voter model is critical \cite{drouffe2}. 
The slow peaking of 
$f_{M}$ for both models is the signature of criticality.

To summarize at this point, there is a strong analogy between the
persistence of luck for the random walk, and phase persistence for
self-similar coarsening systems. However, for the latter, both the
first-passage (or persistence) exponent $\theta $ and the limiting
distributions $\lim_{t\rightarrow \infty }$ $f_{t^{-1}T^{\pm }}(x)$ and 
$f_{M}(x)$ are difficult to determine analytically.

Therefore, instead of considering spatially extended coarsening systems, we
adopt the strategy of investigating simpler (zero-dimensional) stochastic
processes, where the persistence exponent is a parameter in the definition
of the model, yet where the distribution of the occupation time, or of the
magnetization, is non-trivial. So doing we shall gain a better understanding
of the nature of the persistent events encoded in these distributions, and
shall be better prepared to investigate the statistics of the occupation
time for the more difficult statistical mechanical models.

The present work is devoted to the study of a first example of this category
of problems, where the stochastic process $\sigma_{t}$ is generated by a
renewal process, defined as follows. Events occur at the random epochs of
time $t_{1},t_{2},\ldots ,$ from some time origin $t=0$. These events are
considered as the zero crossings of the stochastic process 
$\sigma_{t}=\pm1 $. 
We take the origin of time on a zero crossing. 
This process is known as
a \textit{point process}. When the intervals of time between events, $\tau
_{1}=t_{1},\tau _{2}=t_{2}-t_{1},\ldots $, are independent and identically
distributed random variables with density $\rho (\tau )$, the process thus
formed is a \textit{renewal process}. Hereafter we shall use indifferently
the denominations: events, zero crossings or renewals.

In this model the persistence probability $p_{0}(t)$, that is the
probability that no event occurred up to time $t$, is simply given by the
tail probability: 
\begin{equation}
p_{0}(t)=\mathcal{P}(\tau >t)=\int_{t}^{\infty }d\tau \,\rho (\tau ).  
\label{tail}
\end{equation}

The approach we use is systematic and applies to any distribution $\rho $.
In what follows $\rho $ will be either a narrow distribution with finite
moments, in which case the decay of $p_{0}(t)$, as $t\rightarrow \infty $,
is faster than any power law, or a broad distribution characterized by a
power-law fall-off with index $\theta $: 
\begin{equation}
\int_{t}^{\infty }d\tau \,\rho (\tau )\approx \left( \frac{\tau _{0}}{t}%
\right) ^{\theta }\qquad (0<\theta <2),  \label{def_ro}
\end{equation}
where $\tau _{0}$ is a microscopic time scale. If $\theta <1$ all moments of 
$\rho $ are divergent, while if $1<\theta <2$, the first moment $%
\left\langle \tau \right\rangle $ is finite but higher moments are
divergent. 
In Laplace space, where $s$ is conjugate to $\tau $, for a narrow
distribution we have 
\begin{equation}
\lap{\tau}\rho (\tau )=\hat{\rho}(s)
\comport{=}{ s\rightarrow 0}1-\left\langle 
\tau \right\rangle s+\frac{1}{2}
\left\langle \tau ^{2}\right\rangle s^{2}+\cdots   \label{ro_narrow}
\end{equation}
For a broad distribution, (\ref{def_ro}) yields 
\begin{equation}
\hat{\rho}(s)\comport{\approx }{s\rightarrow 0}\left\{ 
\begin{array}{ll}
1-a\,s^{\theta } & (\theta <1) \\ 
1-\left\langle \tau \right\rangle s+a\,s^{\theta } & (1<\theta <2),
\end{array}
\right. \qquad   \label{ro_broad}
\end{equation}
with $a=|\Gamma (1-\theta )|\tau _{0}^{\theta }$.

The case $\theta =\frac{1}{2}$ accounts for Brownian motion. Indeed, as is
well known, the distribution of first-passage times by the origin behaves at
large times as $\tau ^{-\frac{3}{2}}$, i.e., is in the basin of attraction
of a L\'{e}vy law of index $\frac{1}{2}$ \cite{feller}. This can be simply
worked out in the case of the random walk, where the discreteness of time
allows a natural regularization of the process at short times. Hence, since
we are interested in universal asymptotic properties of the distribution of
the mean magnetization, the description of Brownian motion by a renewal
process with distribution of intervals given by (\ref{def_ro}) or (\ref
{ro_broad}), with index $\theta =\frac{1}{2}$, is faithful.

Finally, let us give the content of the present work and stress its
originality.

Though the study of renewal processes is classical \cite{feller,cox,cox-miller},
fewer references are devoted to renewal processes with broad distributions
of intervals. 
Elements on this question can be found in refs. \cite{feller,dynkin}. 
The present work provides a systematic account of the
theory. In particular we perform the scaling analysis of the distributions
of the various random variables naturally occurring in a renewal process,
such as the number of events between $0$ and $t$, the epoch of the last
event before $t$, the backward and forward times. We also analyze the aging
(i.e., non-stationary) properties of the distribution of the number of
events occurring between two arbitrary instants of time and of the two-time
correlation function, for the case of a broad distribution $\rho $ with $%
\theta <1$.

The study of the limiting distribution of the occupation time of renewal
processes is the subject of the work of Lamperti \cite{lamperti}. 
His main result is the expression of this distribution 
for a broad law of intervals $\rho $, with $\theta <1$ 
(see equation (\ref{lamperti}) below). 
This result
is recovered by a simple method in ref. \cite{baldassarri}, for the case
where $\rho $ is a stable L\'{e}vy distribution of intervals, with $\theta
<1 $. 
However neither the work of Lamperti, nor ref. \cite{baldassarri}, contain 
the analysis of the
scaling of this quantity for the case $1<\theta <2$. 
The present work fills this gap.

Finally, the methods used in the present work provide a firm basis to the
investigation of the distribution of the occupation time for the process
with time-dependent noise considered in \cite{dhar-majumdar}. This model is
a second example of the category of problems mentioned above. There again,
the existence of a renewal process in the model plays a crucial role.\ This
will be the subject of a joint paper \cite{ustocome}. Both examples, the
renewal processes considered in the present work, and that just mentioned,
are deformations of the binomial random walk, or in continuous time, of
Brownian motion. However they lead to non-trivial limiting distributions of
the occupation time and of related quantities, as $t\rightarrow \infty $.

\section{Observables of interest}

\label{observables}

Let us introduce the quantities, the distributions of which will be computed
in the following sections.

First, the number of events which occurred between $0$ and $t$, denoted by $%
N_{t}$, is the random variable for the largest $n$ for which $t_{n}\le t.$
The time of occurrence of the last event before $t$, that is of the $N_{t}$%
-th event, is therefore\footnote{We drop the time dependence of the random
variable when it is in subscript.}
\[
t_{N}=\tau _{1}+\cdots +\tau _{N}. 
\]

The backward recurrence time $B_{t}$ is defined as the length of time
measured backwards from $t$ to the last event before $t$, i.e., 
\[
B_{t}=t-t_{N}, 
\]
while the forward recurrence time (or excess time) $E_{t}$ is the time
interval between $t$ and the next event, 
\[
E_{t}=t_{N+1}-t. 
\]

The occupation times $T_{t}^{+}$ and $T_{t}^{-}$, i.e., the lengths of time
spent by the $\sigma $-process, respectively in the $+$ and $-$ states, up
to time $t$, were defined in the introduction as 
\[
T_{t}^{\pm }=\int_{0}^{t}dt'\frac{1\pm \sigma_{t'}}{2},
\]
hence $t=T_{t}^{+}+T_{t}^{-}$. 
They are simply related to the sum $S_t$ by 
\[
S_{t}=\int_{0}^{t}dt'\sigma_{t'}
=T_{t}^{+}-T_{t}^{-}
=2 T_{t}^{+}-t=t-2 T_{t}^{-}. 
\]
Assume that $\sigma_{t=0}=+1$. Then 
\begin{equation}
\left. 
\begin{array}{l}
T_{t}^{+}=\tau _{1}+\tau _{3}+\cdots +\tau _{N} \\ 
T_{t}^{-}=\tau _{2}+\tau _{4}+\cdots +\tau _{N-1}+B_{t}
\end{array}
\right\} \hbox{if }N_{t}=2k+1 \hbox{ (i.e., }\sigma_{t}=-1)
\label{impair}
\end{equation}
and 
\begin{equation}
\left. 
\begin{array}{l}
T_{t}^{+}=\tau _{1}+\tau _{3}+\cdots +\tau _{N-1}+B_{t} \\ 
T_{t}^{-}=\tau _{2}+\tau _{4}+\cdots +\tau _{N}
\end{array}
\right\} \hbox{if }N_{t}=2k\hbox{ (i.e., }\sigma_{t}=+1).
\label{pair}
\end{equation}
Assume now that $\sigma_{t=0}=-1$. Then, with obvious notations, the
following relation holds 
\begin{equation}
T_{t}^{\pm }(\sigma_{t=0}=-1)=T_{t}^{\mp }(\sigma_{t=0}=+1),  \label{symT}
\end{equation}
hence 
\begin{equation}
S_{t}(\sigma_{t=0}=-1)=-S_{t}(\sigma_{t=0}=+1).  
\label{symM}
\end{equation}

Finally we shall also be interested in two-time quantities, namely the
number of zero crossings which occurred between $t$ and $t+t'$,
given by $N(t,t+t')=N_{t+t'}-N_{t}$, 
and the two-time autocorrelation of the process $\sigma_{t}$, defined as 
$C(t,t+t')=\left\langle\sigma_{t}\sigma_{t+t'}\right\rangle$.

\section{Number of renewals between 0 and $t$}

The probability distribution of the number of events $N_{t}$ between $0$ and 
$t$ reads 
\begin{equation}
p_{n}(t)=\mathcal{P}\left( N_{t}=n\right) =\mathcal{P}\left(
t_{n}<t<t_{n+1}\right) =\left\langle I(t_{n}<t<t_{n+1})\right\rangle \quad
(n\geq 0),  \label{pnt}
\end{equation}
where $I(t_{n}<t<t_{n+1})=1$ if the event inside the parenthesis occurs, and 
$0$ if not. 
Note that $t_{0}=0$. The brackets denotes the average over $\tau
_{1},\tau _{2},\ldots $ The case $n=0$ is accounted for by equation (\ref
{tail}).

Laplace transforming equation (\ref{pnt}) with respect to $t$ yields 
\begin{equation}
\lap{t} p_{n}(t)=
\hat{p}_{n}(s)=
\left\langle
\int_{t_{n}}^{t_{n+1}}dt\, \e^{-st}\right\rangle 
=\left\langle \hbox{e}%
^{-st_{n}}\frac{1-\hbox{e}^{-s\tau _{n+1}}}{s}\right\rangle \qquad (n\geq 0),
\label{pns0}
\end{equation}
and therefore 
\begin{equation}
\hat{p}_{n}(s)=\hat{\rho}(s)^{n}\frac{1-\hat{\rho}(s)}{s}\qquad (n\geq 0).
\label{pns}
\end{equation}
This distribution is normalized since $\sum_{n=0}^{\infty }\hat{p}%
_{n}(s)=1/s $.

From (\ref{pns}) one can easily obtain the moments of $N_{t}$ in Laplace
space. For instance 
\begin{eqnarray}
\lap{t}\langle N_{t}\rangle &=&\sum_{n=1}^{\infty }n\hat{%
p}_{n}(s)=\frac{\hat{\rho}(s)}{s\left( 1-\hat{\rho}(s)\right) },
\label{Naverage} \\
\lap{t}\langle N_{t}^{2}\rangle &=&\sum_{n=1}^{\infty
}n^{2}\hat{p}_{n}(s)=\frac{\hat{\rho}(s)\left( 1+\hat{\rho}(s)\right) }{%
s\left( 1-\hat{\rho}(s)\right) ^{2}}.  \label{N2average}
\end{eqnarray}

We now discuss the above results according to the nature of the distribution
of intervals~$\rho(\tau)$.

\medskip\noindent \textbf{(i)} \textit{Narrow distributions of intervals}

Expanding (\ref{Naverage}) and (\ref{N2average}) as series in $s$, and
performing the inverse Laplace transform term by term, yields 
\begin{eqnarray*}
&&\langle N_{t}\rangle \comport{\approx }{t\rightarrow \infty }\frac{t}{%
\langle \tau \rangle }+c_{1}, \\
&&\langle N_{t}^{2}\rangle -\langle N_{t}\rangle ^{2}
\comport{\approx }{t\rightarrow \infty }\frac{\langle \tau ^{2}
\rangle -\langle \tau
\rangle ^{2}}{\langle \tau \rangle ^{3}}t+c_{2},
\end{eqnarray*}
where the constants $c_1$ and $c_{2}$ can be expressed in terms of
the moments of $\rho $. 
More generally, all the cumulants of $N_{t}$ scale
as $t$, as would be the case for the sum of $t$ independent random
variables, and therefore $N_{t}$ obeys the central limit theorem \cite
{cox,cox-miller}.

\medskip \noindent \textbf{(ii)} 
\textit{Broad distributions of intervals with index }%
$\theta <1$

Using equation (\ref{ro_broad}), (\ref{Naverage}) yields 
\begin{equation}
\langle N_{t}\rangle \comport{\approx }{t\rightarrow \infty }\frac{\sin
\pi \theta }{\pi \theta }\,\left( \frac{t}{\tau _{0}}\right) ^{\theta },
\label{Nmean}
\end{equation}
while (\ref{N2average}) yields $\langle N_{t}^{2}\rangle \sim t^{2\theta }$.
More generally one expects that the cumulant of order $k$ of $N_{t}$ scales
as $t^{k\theta }$. 
Indeed, setting
\[
N_{t}=\frac{(t/\tau _{0})^{\theta }}{\Gamma (1-\theta )}X_{t}, 
\]
where we keep the same notation $X_{t}$ for the scaling variable, we obtain,
using (\ref{ro_broad}) and  
\begin{equation}
p_{n}(t)=\int \frac{ds}{2\pi \mathrm{i}}\hbox{e}^{st}\hat{\rho}(s)^{n}%
\frac{1-\hat{\rho}(s)}{s},  
\label{pn}
\end{equation}
the limiting distribution of 
$X_{t}$,\thinspace 
as $t\rightarrow \infty $, 
\begin{equation}
f_{X}(x)
=\int \frac{dz}{2\pi {\i}}z^{\theta -1}\hbox{e}^{z-xz^{\theta
}}\qquad (0<x<\infty).  
\label{fX}
\end{equation}
For small $x$, expanding the integrand in the right side and folding the
contour around the negative real axis yields 
\[
f_{X}(x)\comport{=}{x\rightarrow 0}\frac{1}{\Gamma (1-\theta )}-\frac{x}{%
\Gamma (1-2\theta )}+\cdots 
\]
At large $x$, applying the steepest-descent method, we find the stretched
exponential fall-off 
\[
f_{X}(x)\comport{\sim }{x\rightarrow \infty }\exp \left( -(1-\theta
)\left( \theta ^{\theta }x\right) ^{1/(1-\theta )}\right) , 
\]
which demonstrates that all the moments of $X$ are finite.

For $\theta =1/2$, the distribution of $X$ is given by a half Gaussian: 
\[
f_{X}(x)=\frac{\hbox{e}^{-x^{2}/4}}{\sqrt{\pi }}\qquad (0<x<\infty). 
\]

\medskip \noindent \textbf{(iii)} 
\textit{Broad distributions of intervals with index }$1<\theta <2$

Using equation (\ref{ro_broad}), we obtain 
\begin{eqnarray}
&&\langle N_{t}\rangle \comport{\approx }{t\rightarrow \infty }\frac{t}
{\langle \tau \rangle }+\frac{\tau _{0}^{\theta }}{(\theta -1)(2-\theta
)\langle \tau \rangle ^{2}}t^{2-\theta },  \label{Nt} \\
&&\langle N_{t}^{2}\rangle -\langle N_{t}\rangle ^{2}
\comport{\approx }{t\rightarrow \infty }\frac{2\tau _{0}^{\theta }}{(2-\theta
)(3-\theta )\langle \tau \rangle ^{3}}t^{3-\theta }.  \label{N2t}
\end{eqnarray}
The fluctuations of the variable $N_{t}$ around its mean $t/\left\langle
\tau \right\rangle $ are therefore no longer characterized by a single
scale. These fluctuations are very large as we now show.

We set, still keeping the same notation $X_{t}$ for the scaling variable, 
\begin{equation}
N_{t}=\frac{t}{\left\langle \tau \right\rangle }+\frac{\tau _{0}}{%
\left\langle \tau \right\rangle }\left( -\Gamma (1-\theta )\frac{t}{%
\left\langle \tau \right\rangle }\right) ^{1/\theta }X_{t}.  \label{defY}
\end{equation}
Then, using equations~(\ref{ro_broad}) and~(\ref{pn}), we obtain 
the limiting distribution of $X_{t}$,
as $t\rightarrow \infty $,  
\begin{equation}
f_{X}(x)=\int \frac{dz}{2\pi {\i}}\,\hbox{e}^{-zx+z^{\theta }}.
\label{levyY}
\end{equation}
For large positive values of $x$, $f_{X}$ falls off exponentially as 
\[
f_{X}(x)\comport{\sim }{x\rightarrow +\infty }\exp \left( -(\theta
-1)\left( \frac{x}{\theta }\right) ^{\theta /(\theta -1)}\right) . 
\]
For large negative values of $x$, linearizing the integrand of equation~(\ref
{levyY}) with respect to $z^{\theta }$, and folding the contour around the
negative real axis, we find 
\[
f_{X}(x)\comport{\approx }{x\rightarrow -\infty }\frac{|x|^{-\theta -1}}{%
\Gamma (-\theta )}. 
\]
As a consequence, $\left\langle X_{t}\right\rangle $ is finite, and all
higher order moments diverge. 
Actually $\left\langle X_{t}\right\rangle $
vanishes, as seen from (\ref{Nt}) and (\ref{defY}), because the difference
of exponents $2-\theta -1/\theta =-(\theta -1)^{2}/\theta $ is negative. 
One can also check on~(\ref{N2t}) that 
$\left\langle X_{t}^{2}\right\rangle $
is divergent, since the difference of exponents 
$3-\theta -2/\theta =(\theta-1)(2-\theta )/\theta $ is now positive.

In the limit $\theta \to 2$, $f_X$ is a Gaussian: 
\[
f_{X}(x)\comport{\rightarrow }{\theta \rightarrow 2}\frac{\hbox{e}%
^{-x^{2}/4}}{2\sqrt{\pi }}. 
\]
Related considerations can be found in ref. \cite{feller}, vol. 2, p. 373.

\section{Epoch of last renewal $t_{N}$}

\label{tN}

We begin our study of the distributions of the random variables $t_{N}$, 
$B_{t}$, $E_{t}$, $T_{t}^{\pm }$, and $S_{t}$ appearing in section \ref
{observables} by the example of $t_{N}$, the epoch of the last renewal
before~$t$.

This quantity is the sum of a random number $N_{t}$ of random variables $%
\tau _{1},\tau _{2},\ldots $, i.e., it is a function of the joint random
variables $\{\tau _{1},\tau _{2},\ldots ,\tau _{N}\}$. Now, the latter are 
\textit{not} independent, although the intervals $\tau _{1},\tau _{2},\ldots 
$ are, by definition, independent. This is due to the very definition of $%
N_{t}$. Indeed if, say, $N_{t}$ takes the value $n,$ then the sum $%
t_{n}=\sum_{i=1}^{n}\tau _{i}$ is constrained to be less than $t$.
Therefore, in particular, each individual interval $\tau _{i}$ is
constrained to be less than $t$.

These considerations will be now made more precise, by the computation of
the distribution of $t_{N}$, below, and by that of the joint probability
density function of $\{\tau _{1},\tau _{2},\ldots ,\tau _{N}\}$ and $N_{t}$,
in the next section.

The joint probability distribution $f_{t_{N},N}$ of the random variables $t_{N}$
and $N_{t}$ reads 
\[
f_{t_{N},N}(t;y,n)=\frac{d}{d y}\mathcal{P}\left(
t_{N}<y,N_{t}=n\right) =\left\langle \delta
(y-t_{N})\,I(t_{n}<t<t_{n+1})\right\rangle ,
\]
from which one deduces the density $f_{t_{N}}$ of $t_{N}$%
\[
f_{t_{N}}(t;y)=\frac{d}{d y}\mathcal{P}\left( t_{N}<y\right)
=\left\langle \delta (y-t_{N})\right\rangle =\sum_{n=0}^{\infty
}f_{t_{N},N}(t;y,n). 
\]
In Laplace space, where $s$ is conjugate to $t$ and $u$ to $y$, 
\begin{eqnarray}
\lap{t,y} f_{t_{N},N}(t;y,n) &=&\hat{f}%
_{t_{N},N}(s;u,n)=\left\langle \hbox{e}^{-ut_{n}}\int_{t_{n}}^{t_{n+1}}dt\,%
\hbox{e}^{-st}\;\right\rangle  \nonumber \\
&=&\hat{\rho}(s+u)^{n}\frac{1-\hat{\rho}(s)}{s}\qquad (n\geq 0).
\label{ftNn}
\end{eqnarray}
The distribution (\ref{pns}) of $N_{t}$ is recovered by setting $u=0$ in (%
\ref{ftNn}). Summing over $n$ gives the distribution of $t_{N}$, in Laplace
space: 
\begin{equation}
\lap{t}\left\langle \hbox{e}^{-ut_{N}}\right\rangle =%
\hat{f}_{t_{N}}(s;u)=\frac{1}{1-\hat{\rho}(s+u)}\frac{1-\hat{\rho}(s)}{s},
\label{ftN}
\end{equation}
which is normalized since $\hat{f}_{t_{N}}(s;u=0)=1/s$.

The case where the distribution of intervals $\rho(\tau)$ is broad, with 
$\theta <1$ (see equation (\ref{ro_broad})), is of particular interest since
it leads to a limiting distribution for the random variable $t_{N}/t$.

In the long-time scaling regime, where $t$ and $t_{N}$ are both large and
comparable, or $u,s$ small and comparable, we get 
\[
\hat{f}_{t_{N}}(s;u)\approx s^{\theta -1}\,(s+u)^{-\theta }. 
\]
This yields, using the method of appendix~B, the limiting distribution for
the rescaled variable $t^{-1}t_{N}$, as $t\rightarrow \infty $, 
\begin{equation}
\lim_{t\rightarrow \infty }f_{t^{-1}t_{N}}(x)=\frac{\sin \pi \theta }{\pi }%
x^{\theta -1}\,(1-x)^{-\theta }=\beta_{\theta ,1-\theta }(x)\qquad (0<x<1),
\label{tNlimit}
\end{equation}
with $x=y/t$, and where 
\[
\beta _{a,b}(x)=\frac{\Gamma (a+b)}{\Gamma (a)\Gamma (b)}x^{a-1}(1-x)^{b-1} 
\]
is the beta distribution on $[0,1]$. As a consequence 
\begin{equation}
\left\langle t_{N}\right\rangle \comport{\approx }{t\rightarrow \infty }
\theta t.  \label{tNmean}
\end{equation}
In the particular case $\theta =\frac{1}{2}$, we have 
\begin{equation}
\lim_{t\rightarrow \infty }f_{t^{-1}t_{N}}(x)
=\frac{1}{\pi\sqrt{x(1-x)}},
\end{equation}
which is the arcsine law on $[0,1]$. 
This is a well-known property of Brownian motion~\cite{feller}.

\section{Interdependence of $\{\tau _{1},\ldots ,\tau _{N}\}$}

\label{section_joint}

The purpose of this section is to 
identify the interdependence of the $N_{t}$ first time intervals 
$\{\tau _{1}$, $\tau _{2}$, $\ldots $, $\tau _{N}\}$,
a point both of conceptual and of practical importance
(see e.g.~\cite{ustocome}). 

In order to do so,  we compute the joint
probability distribution of these random variables and of $N_{t}$, denoted by 
\[
f_{\{\tau _{1},\tau _{2},\ldots ,\tau _{N}\},N}(t;\tau _{1},\tau _{2},\ldots
,\tau _{n},n). 
\]
Generalizing the calculations done above, one has, in the Laplace space of
all temporal variables, 
\[
\hat{f}_{\{\tau _{1},\tau _{2},\ldots ,\tau _{N}\},N}(s;u_{1},\ldots
,u_{n},n)=\left\langle 
\prod_{i=1}^{n}\hbox{e}^{-u_{i}\tau_{i}}
\int_{t_{n}}^{t_{n+1}}dt\,\hbox{e}^{-st}\right\rangle ,
\]
resulting in 
\[
\hat{f}_{\{\tau _{1},\tau _{2},\ldots ,\tau _{N}\},N}(s;u_{1},u_{2},\ldots
,u_{n},n)=\frac{1-\hat{\rho}(s)}{s}\prod_{i=1}^{n}\hat{\rho}(s+u_{i})\qquad
(n\geq 0), 
\]
where the empty product is equal to 1 for $n=0$.

The marginal distribution (\ref{pns}) of $N_{t}$ can be recovered by
setting all the $u_{i}=0$ in the above expression. By inversion with respect
to the variables $\left\{ u_{i}\right\} $, one gets 
\begin{equation}
\hat{f}_{\{\tau _{1},\tau _{2},\ldots ,\tau _{N}\},N}(s;\tau _{1},\tau
_{2},\ldots ,\tau _{n},n)=\frac{1-\hat{\rho}(s)}{s}\prod_{i=1}^{n}\rho (\tau
_{i})\hbox{e}^{-s\tau _{i}}  \label{joint}
\end{equation}
and finally 
\[
f_{\{\tau _{1},\tau _{2},\ldots ,\tau _{N}\},N}(t;\tau _{1},\tau _{2},\ldots
,\tau _{n},n)=\left( \prod_{i=1}^{n}\rho (\tau _{i})\right) \mathcal{P}%
(N_{t-t_{n}}=0)\,\Theta (t-t_{n}), 
\]
where $\Theta (x)$ is Heaviside step function. This expression clearly
exhibits the interdependence of the random variables $\{\tau _{1},\tau
_{2},\ldots ,\tau _{N}\}$.

Let us investigate the distribution of any one of the $\tau _{i}$, say $\tau
_{1}$. We denote this random variable by $\tau _{t}$ in order to enhance the
fact that its distribution is constrained. By integration of (\ref{joint})
on $\tau _{2},\tau _{3},\ldots ,\tau _{n}$, we obtain the joint distribution
of $N_{t}$ and $\tau _{t}$ in Laplace space: 
\[
\hat{f}_{\tau _{t},N}(s;\tau ,n)=\frac{1-\hat{\rho}(s)}{s}\times \left\{ 
\begin{array}{ll}
\hat{\rho}(s)^{n-1}\hbox{e}^{-s\tau }\rho (\tau ) & (n\geq 1), \\ 
\delta (\tau ) & (n=0),
\end{array}
\right. 
\]
from which one deduces, by summation upon $n$, that 
\begin{equation}
\hat{f}_{\tau _{t}}(s;\tau )=\rho (\tau )\frac{\hbox{e}^{-s\tau }}{s}+\delta
(\tau )\frac{1-\hat{\rho}(s)}{s}.  \label{ftaus}
\end{equation}
This can be finally inverted, yielding 
\begin{equation}
f_{\tau _{t}}(t;\tau )=\rho (\tau )\Theta (t-\tau )+\delta (\tau )\,p_{0}(t).
\label{ftau}
\end{equation}
This result can be interpreted as follows. The a priori distribution $\rho
(\tau )$ is unaffected as long as $\tau $ is less than the observation time $%
t$. The complementary event, which has probability $p_{0}(t)$, corresponds
formally to $\tau =0$.

As a consequence of (\ref{ftaus}) and (\ref{ftau}) we have 
\begin{equation}
\lap{t}\left\langle \tau _{t}\right\rangle =\frac{1}{s}%
\int_{0}^{\infty }d\tau \,\tau \,\rho (\tau )\hbox{e}^{-s\tau }=-\frac{1}{s}%
\frac{d\hat{\rho}(s)}{ds},  \label{taumeans}
\end{equation}
and 
\begin{equation}
\left\langle \tau _{t}\right\rangle =\int_{0}^{\infty }d\tau \,\tau
\,f_{\tau _{t}}(t;\tau )=\int_{0}^{t}d\tau \,\tau \,\rho (\tau
)=\left\langle \tau \right\rangle -\int_{t}^{\infty }d\tau \,\tau \,\rho
(\tau ),  \label{taumean}
\end{equation}
where the last expression holds when the a priori average $\left\langle \tau
\right\rangle $ is finite.

Let us discuss the above results according to the nature of the distribution
of intervals~$\rho(\tau)$.

\medskip \noindent \textbf{(i)} \textit{Narrow distributions of intervals}

If the distribution $\rho $ is narrow, then  (\ref{taumean}) 
shows that  $\langle\tau_t\rangle$ converges to $\langle\tau\rangle$
very rapidly.
For instance, if $\rho $ is exponential, then $\left\langle \tau
_{t}\right\rangle =[1-(1+\lambda t)$e$^{-\lambda t}]/\lambda $:
the decay of $\langle\tau\rangle-\langle\tau_t\rangle$, 
with $\langle\tau\rangle=1/\lambda$, is exponential.

\medskip \noindent \textbf{(ii)} \textit{Broad distributions of intervals with index }%
$\theta <1$

If the distribution $\rho $ is broad, with $\theta <1$, then from (\ref
{taumeans}) 
\[
\left\langle \tau _{t}\right\rangle 
\comport{\approx }{t\rightarrow \infty }
\frac{\theta \tau _{0}^{\theta }}{1-\theta }t^{1-\theta }. 
\]
The interpretation of this last result is that $t_{N}\sim t$ is
the sum of $N_{t}\sim t^{\theta }$ time intervals 
$\tau _{t}\sim t^{1-\theta}$. 
However $\left\langle t_{N}\right\rangle \equiv \left\langle N_{t}\tau
_{t}\right\rangle \neq \left\langle N_{t}\right\rangle \left\langle \tau
_{t}\right\rangle $ (see equations (\ref{Nmean}) and (\ref{tNmean})).

\medskip \noindent \textbf{(iii)} \textit{Broad distributions of intervals with index 
}$1<\theta <2$

We now obtain 
\[
\left\langle \tau _{t}\right\rangle 
\comport{\approx }{t\rightarrow \infty }\left\langle \tau \right\rangle -\frac{\theta \tau _{0}^{\theta }}{%
\theta -1}t^{-(\theta -1)}, 
\]
showing that $\left\langle \tau _{t}\right\rangle $ converges to $%
\left\langle \tau \right\rangle $ very slowly.

\section{Backward and forward recurrence times}

\label{section_back}

The distributions of $B_{t}=t-t_{N}$ and $E_{t}=t_{N+1}-t$ can be obtained
using the methods of the previous sections. We have 
\begin{eqnarray*}
f_{B,N}(t;y,n) &=&\left\langle \delta
(y-t+t_{n})\,I(t_{n}<t<t_{n+1})\right\rangle , \\
f_{E,N}(t;y,n) &=&\left\langle \delta
(y-t_{n+1}+t)\,I(t_{n}<t<t_{n+1})\right\rangle .
\end{eqnarray*}
In Laplace space, where $s$ is conjugate to $t$ and $u$ to $y$, 
\begin{eqnarray*}
\lap{t,y}f_{B,N}(t;y,n) &=&\hat{f}_{B,N}(s;u,n)=\left%
\langle \int_{t_{n}}^{t_{n+1}}dt\,\hbox{e}^{-st}\,\hbox{e}%
^{-u(t-t_{n})}\right\rangle , \\
\lap{t,y}f_{E,N}(t;y,n) &=&\hat{f}_{E,N}(s;u,n)=\left%
\langle \int_{t_{n}}^{t_{n+1}}dt\,\hbox{e}^{-st}\hbox{e}^{-u\left(
t_{n+1}-t\right) }\right\rangle ,
\end{eqnarray*}
hence, for $n\geq 0$, 
\begin{eqnarray*}
\hat{f}_{B,N}(s;u,n) &=&\hat{\rho}(s)^{n}\frac{1-\hat{\rho}(s+u)}{s+u}, \\
\hat{f}_{E,N}(s;u,n) &=&\hat{\rho}(s)^{n}\frac{\hat{\rho}(s)-\hat{\rho}(u)}{%
u-s},
\end{eqnarray*}
and therefore 
\begin{eqnarray}
\hat{f}_{B}(s;u) &=&\frac{1-\hat{\rho}(s+u)}{s+u}\frac{1}{1-\hat{\rho}(s)},
\label{fB} \\
\hat{f}_{E}(s;u) &=&\frac{\hat{\rho}(u)-\hat{\rho}(s)}{s-u}\frac{1}{1-\hat{%
\rho}(s)}.  \label{fE}
\end{eqnarray}
Equations (\ref{fB}) and (\ref{ftN}) are related by $\hat{f}_{B}(s;u)=\hat{f}%
_{t_{N}}(s+u;-u)$, expressing the identity $t_{N}+B_{t}=t$.

We now discuss the above results according to the nature of the distribution 
$\rho(\tau)$.

\medskip \noindent \textbf{(i)} \textit{Narrow distributions of intervals}

For distributions with finite moments, equilibrium is attained at long
times, for both the backward and forward recurrence times, with a common
distribution given in Laplace space by 
\begin{equation}
\hat{f}_{B,{\eq}}(u)
=\hat{f}_{E,{\eq}}(u)
=\lim_{s\rightarrow 0}s\hat{f}_{B}(s;u)
=\lim_{s\rightarrow 0}s\hat{f}_{E}(s;u)
=\frac{1-\hat{\rho}(u)}{\left\langle\tau \right\rangle u}.  
\label{fBEeq1}
\end{equation}
By inversion we obtain 
\begin{equation}
f_{B,{\eq}}(y)=f_{E,{\eq}}(y)=\frac{1}{\left\langle \tau
\right\rangle }\int_{y}^{\infty }\,d\tau \,\rho (\tau )=\frac{p_{0}(y)}{%
\left\langle \tau \right\rangle }\,.  
\label{f_eq}
\end{equation}

In the particular case where $\rho $ is exponential, inversion of (\ref{fB})
yields, for finite $t$, 
\[
f_{B}(t;y)=\lambda \hbox{e}^{-\lambda y}\Theta (t-y)+\hbox{e}^{-\lambda
t}\delta (t-y). 
\]
The weight of the second term is simply $\mathcal{P}(B_{t}=t)=p_{0}(t)$.
Similarly, by inversion of (\ref{fE}), we have 
\[
f_{E}(t;y)=f_{E,{\eq}}(y)=f_{B,{\eq}}(y)=\lambda \hbox{e}^{-\lambda
y}=\rho (y). 
\]

\medskip \noindent \textbf{(ii)} 
\textit{Broad distributions of intervals with index }%
$\theta <1$

We obtain, in the long-time scaling regime, i.e., for $u,s$ small and
comparable, 
\[
\hat{f}_{B}(s;u)\approx s^{-\theta }(s+u)^{\theta -1}, 
\]
which, by inversion, using the method of appendix~B, yields the limiting
distribution 
\begin{equation}
\lim_{t\rightarrow \infty }f_{t^{-1}B}(x)=\frac{\sin \pi \theta }{\pi }%
x^{-\theta }(1-x)^{\theta -1}=\beta _{1-\theta ,\theta }(x)\qquad (0<x<1).
\label{Blimit}
\end{equation}
This result is consistent with (\ref{tNlimit}). Similarly, 
\[
\hat{f}_{E}(s;u)\approx \frac{u^{\theta }-s^{\theta }}{s^{\theta }(u-s)}, 
\]
yielding 
\begin{equation}
\lim_{t\rightarrow \infty }f_{t^{-1}E}(x)=\frac{\sin \pi \theta }{\pi }\frac{%
1}{x^{\theta }(1+x)}\qquad (0<x<\infty ).  
\label{fEscal}
\end{equation}
Let us point out that the limiting distribution of $t/t_{N+1}$, as $%
t\rightarrow \infty $, is given by (\ref{tNlimit}).

For $\theta =\frac{1}{2}$, we obtain 
\[
\lim_{t\rightarrow \infty }f_{t^{-1}B}(x)=\frac{1}{\pi \sqrt{x(1-x)}}\qquad
(0<x<1), 
\]
which is the arcsine law on $[0,1]$. Similarly 
\[
\lim_{t\rightarrow \infty }f_{t^{-1}E}(x)=\frac{1}{\pi (1+x)\sqrt{x}}\qquad
(0<x<\infty ). 
\]

\medskip \noindent \textbf{(iii)} \textit{Broad distributions of intervals with index 
}$1<\theta <2$

The backward and forward recurrence times still admit the common limiting
distribution~(\ref{f_eq}), which in the present case has the following
asymptotic behavior 
\begin{equation}
f_{B,{\eq}}(y)=f_{E,{\eq}}(y)
\comport{\approx }{y\rightarrow \infty }\frac{1}{\left\langle \tau \right\rangle }\left( \frac{\tau _{0}}{y}%
\right) ^{\theta },  \label{fBEeq}
\end{equation}
characteristic of a broad distribution of index $\theta -1<1$. In other
words, at equilibrium the average backward and forward recurrence times
diverge. However their long-time behavior can be computed from (\ref{fB})
and (\ref{fE}), yielding 
\[
\lap{t}\left\langle B_{t}\right\rangle =\frac{1-\hat{\rho%
}(s)+s\,d\hat{\rho}(s)/ds}{s^{2}(1-\hat{\rho}(s))},\qquad 
\lap{t}\left\langle E_{t}\right\rangle =\frac{\hat{\rho}%
(s)-1+\left\langle \tau \right\rangle s}{s^{2}(1-\hat{\rho}(s))}, 
\]
from which it follows that 
\[
\left\langle B_{t}\right\rangle \comport{\approx }{t\rightarrow \infty }
\frac{\tau _{0}^{\theta }}{(2-\theta )\left\langle \tau \right\rangle }%
t^{2-\theta },\qquad \left\langle E_{t}\right\rangle 
\comport{\approx }{t\rightarrow \infty }\frac{\tau _{0}^{\theta }}{(\theta
-1)(2-\theta )\left\langle \tau \right\rangle }t^{2-\theta }. 
\]

\section{Occupation time and mean magnetization}

\label{occup}

The central investigation of the present work concerns the determination of
the distributions of the occupation times $T_{t}^{\pm }$ and of the sum $S_t$ 
(or of the mean magnetization $M_{t}$). 

Let us denote by $f_{T^{\pm }}^{\sigma_{0}}$ and $f_{S}^{\sigma_{0}}$ the
probability density functions of these quantities for a fixed value of 
$\sigma_{0}\equiv \sigma_{t=0}$, and by $f_{T^{\pm }}$ and $f_{S}$ the
corresponding probability density functions after averaging over 
$\sigma_{0}=\pm 1$ with equal weights: 
\[
f=\frac{1}{2}(f^{+}+f^{-}). 
\]
The symmetry properties (\ref{symT}) and (\ref{symM}) imply that 
\begin{equation}
f_{T^{+}}(t;y)=f_{T^{-}}(t;y),\qquad f_{S}(t;y)=f_{S}(t;-y).  
\label{sym}
\end{equation}

Following the methods used in the previous sections, we define the joint
probability distribution of $T_{t}^{+}$ and $N_{t}$ at fixed $\sigma_{0}$
as 
\[
f_{T^{+},N}^{\sigma_{0}}(t;y,n)=\frac{d}{dy}\mathcal{P}\left(
T_{t}^{+}<y,N_{t}=n\right) =\left\langle \delta \left( y-T_{t}^{+}\right)
I\left( t_{n}\le t<t_{n+1}\right) \right\rangle , 
\]
and 
\[
f_{T^{+}}^{\sigma_{0}}(t;y)=\sum_{n=0}^{\infty }f_{T^{+},N}^{\sigma
_{0}}(t;y,n)=\left\langle \delta \left( y-T_{t}^{+}\right) \right\rangle . 
\]
For $\sigma_{0}=+1$, Laplace transforming with respect to $t$ and $y$,
using equations (\ref{impair}) and (\ref{pair}), we find 
\begin{eqnarray*}
\hat{f}_{T^{+},N}^{+}(s;u,2k+1) &=&\hat{\rho}^{k+1}(s+u)\,\hat{\rho}^{k}(s)%
\frac{1-\hat{\rho}(s)}{s}, \\
\hat{f}_{T^{+},N}^{+}(s;u,2k) &=&\hat{\rho}^{k}(s+u)\,\hat{\rho}^{k}(s)\frac{%
1-\hat{\rho}(s+u)}{s+u},
\end{eqnarray*}
hence, summing over $k,$%
\[
\hat{f}_{T^{+}}^{+}(s;u)=\left( \frac{1-\hat{\rho}(s+u)}{s+u}+\hat{\rho}(s+u)%
\frac{1-\hat{\rho}(s)}{s}\right) \frac{1}{1-\hat{\rho}(s)\hat{\rho}(s+u)}. 
\]

Using the property $T_{t}^{-}=t-T_{t}^{+}$, and (\ref{sym}), we get 
\[
\hat{f}_{T^{+}}(s;u)=\frac{1}{2}\left( \hat{f}_{T^{+}}^{+}(s;u)+\hat{f}%
_{T^{+}}^{+}(s+u;-u)\right) =\hat{f}_{T^{+}}(s+u;-u). 
\]
Similarly, using the property $S_{t}=2T_{t}^{+}-t$, and (\ref{sym}), we get 
\[
\hat{f}_{S}(s;u)=\hat{f}_{S}(s;-u)=\hat{f}_{T^{+}}(s-u;2u). 
\]
(Here $\hat{f}_{S}(t;u)$ is the bilateral Laplace transform with respect to $y$.
See appendix~A for the notation.)

The final results read 
\begin{eqnarray}
\hat{f}_{T^{\pm }}(s;u) &=&\frac{2s\left( 1-\hat{\rho}(s+u)\hat{\rho}%
(s)\right) +u\left( 1+\hat{\rho}(s+u)\right) \left( 1-\hat{\rho}(s)\right) }{%
2s(s+u)\left( 1-\hat{\rho}(s+u)\hat{\rho}(s)\right) },  \nonumber \\
\hat{f}_{S}(s;u) &=&\frac{s\left( 1-\hat{\rho}(s+u)\hat{\rho}(s-u)\right)
+u\left( \hat{\rho}(s+u)-\hat{\rho}(s-u)\right) }{(s^{2}-u^{2})\left( 1-\hat{%
\rho}(s+u)\hat{\rho}(s-u)\right) }.  \label{fS}
\end{eqnarray}

Let us discuss these results according to the nature of the distribution of
intervals~$\rho(\tau)$.

\medskip \noindent \textbf{(i)} \textit{Narrow distributions of intervals}

If the distribution $\rho $ is narrow, implying that the correlations
between the sign process $\sigma_t$ at two instants of time are
short-ranged (see section \ref{auto}), it is intuitively clear that
$S_{t}$ should scale as the sum of $t$
independent random variables and therefore obey the central limit theorem. 
In the context of the present work, we are especially interested in
large deviations, i.e., rare (persistent) events where 
$S_t$ deviates from its mean.
In particular, the probability that $S_t$ is {\it equal} to $t$ is 
identical to $p_0(t)$, the persistence probability.
These points are now made more precise.

First, expanding the right-hand side of (\ref{fS}) 
to second order in $u$, and performing the inverse Laplace transform,
yields 
\begin{equation}
\left\langle S_{t}^{2}\right\rangle 
\comport{\approx }{t\rightarrow \infty }
\frac{\left\langle \tau ^{2}\right\rangle -\left\langle \tau
\right\rangle ^{2}}{\left\langle \tau \right\rangle }t.  
\label{S2_narrow}
\end{equation}
More generally, it can be checked that all the cumulants of $S_t$ scale as $t$. 
Then, 
\begin{equation}
\hat{f}_{S}(t;u)=\left\langle \hbox{e}^{-uS_{t}}\right\rangle 
=\e^{K_S(t;u)}
\comport{\sim }{t\rightarrow \infty }\hbox{e}^{t\Phi (u)},  
\label{large1}
\end{equation}
where $K_S(t;u)$ is the generating function of cumulants
of $S_t$, and $\Phi(u)$  is determined below.
By inversion of (\ref{large1}), 
\[
f_{S}(t;y)\comport{\sim }{t\rightarrow \infty }\int \frac{du}{2{\i}%
\pi }\hbox{e}^{uy+t\Phi (u)}.
\]
In the large-deviation regime, i.e., when $t$ and $y$ are simultaneously large and
$x=y/t$ finite, the saddle-point method yields
\[
f_{M}(t;x)\comport{\sim }{t\rightarrow \infty }\int \frac{du}{2{\i}%
\pi }\hbox{e}^{t\left[ ux+\Phi (u)\right] }
\comport{\sim }{t\rightarrow \infty }\hbox{e}^{-t\Sigma (x)},
\]
where
\[
\Sigma (x)=-\min_{u}\left( ux+\Phi (u)\right)\qquad (-1<x<1) 
\]
is the large-deviation function (or entropy) for $M_{t}$. 
The functions $\Sigma (x)$ and $\Phi (u)$ are mutual Legendre transforms: 
\[
\Sigma (x)+\Phi (u)=-ux,\qquad u=-\frac{d\Sigma }{dx},\qquad x=-\frac{d\Phi 
}{du}.
\]

Equation (\ref{large1}) implies that $\hat{f}_{S}(s;u)$ is singular for $%
s=\Phi (u)$, and therefore, using (\ref{fS}), that 
\begin{equation}
\hat{\rho}\left( \Phi (u)+u\right) \hat{\rho}\left( \Phi (u)-u\right) =1,
\label{eq_phi}
\end{equation}
which determines implicitly $\Phi (u)$. 
For $u\rightarrow 0$, using $\hat{%
\rho}(u)\approx 1-u\left\langle \tau \right\rangle +\frac{1}{2}%
u^{2}\left\langle \tau ^{2}\right\rangle $, 
equation~(\ref{eq_phi}) gives 
\[
\Phi (u)\comport{\approx }{u\rightarrow 0}\frac{\left\langle \tau
^{2}\right\rangle -\left\langle \tau \right\rangle ^{2}}{2\left\langle \tau
\right\rangle }u^{2}, 
\]
hence 
\[
\Sigma (x)\comport{\approx }{x\rightarrow 0}\frac{\left\langle \tau
\right\rangle }{2(\left\langle \tau ^{2}\right\rangle -\left\langle \tau
\right\rangle ^{2})}x^{2}, 
\]
yielding a Gaussian distribution for $S_{t}$, thus recovering the central
limit theorem for this quantity.

In the particular case of an exponential distribution $\rho (\tau )$, all
results become explicit. We have 
\[
\hat{f}_{S}(s;u)=\frac{2\lambda +s}{s^{2}+2\lambda s-u^{2}}, 
\]
and 
\[
\Phi (u)=\sqrt{\lambda ^{2}+u^{2}}-\lambda ,\qquad \Sigma (x)=\lambda \left(
1-\sqrt{1-x^{2}}\right) . 
\]

\medskip \noindent \textbf{(ii)} \textit{Broad distributions of intervals with index }%
$\theta <1$

Using equation (\ref{ro_broad}), we have, in the long-time scaling regime
where $u$ and $s$ are small and comparable, 
\[
\hat{f}_{S}(s;u)\approx \frac{(s+u)^{\theta -1}+\left( s-u\right) ^{\theta
-1}}{(s+u)^{\theta }+\left( s-u\right) ^{\theta }}, 
\]
which, by inversion, using the method of appendix~B, yields, as $%
t\rightarrow \infty $, the limiting distribution for the mean magnetization $%
M_{t}$, 
\begin{equation}
f_{M}(x)=\frac{2\sin \pi \theta }{\pi }\frac{\left( 1-x^{2}\right) ^{\theta
-1}}{(1+x)^{2\theta }+(1-x)^{2\theta }+2\cos \pi \theta \left(
1-x^{2}\right) ^{\theta }}.  \label{fMasympt}
\end{equation}
For $x\to 0$, the expansion 
\[
f_{M}(x)=\frac{2\tan (\pi \theta /2)}{\pi }\left( 1+\frac{\cos ^{2}(\pi
\theta /2)-\theta ^{2}}{\cos ^{2}(\pi \theta /2)}\,x^{2}+\cdots \right) 
\]
shows that $x=0$ is a minimum of $f_{M}(x)$ for $\theta <\theta _{c}$, while
it is a maximum for $\theta >\theta _{c}$, with $\theta _{c}=\cos (\pi
\theta _{c}/2)$, yielding $\theta _{c}=0.594611$ \cite{baldassarri}.

For $x\to \pm 1$, $f_{M}(x)$ diverges as 
\begin{equation}
f_{M}(x)\comport{\approx }{x\rightarrow \pm 1}{\approx }\frac{\sin \pi \theta }{\pi 
}2^{-\theta }(1\mp x)^{\theta -1}.  \label{fM_x=1}
\end{equation}
Comparing the amplitude of this power-law divergence to that of the
symmetric beta distribution over $[-1,1]$ of same index, 
\[
\beta (x)=\frac{\Gamma (\theta +\frac{1}{2})}{\Gamma (\theta )\sqrt{\pi }}%
\,(1-x^{2})^{\theta -1}, 
\]
we have 
\[
\lim_{x\to \pm 1}\frac{f_{M}(x)}{\beta (x)}=B(\theta )=\frac{\Gamma (\theta )%
}{\Gamma (2\theta )\Gamma (1-\theta )}. 
\]
The amplitude ratio $B(\theta )$ decreases from $B(0)=2$ to $B(1)=0$; it is
equal to $1$ for $\theta =\frac{1}{2}$ ~(see~equation~(\ref{arcsin})).

Equation~(\ref{d}) yields the moments of $f_{M}$, 
\begin{equation}
\left\langle M^{2}\right\rangle =1-\theta ,\quad \left\langle
M^{4}\right\rangle =1-\frac{\theta }{3}\,(4-\theta ^{2}),\quad \left\langle
M^{6}\right\rangle =1-\frac{\theta }{15}\,(23-10\theta ^{2}+2\theta ^{4}),
\label{M2_broad1}
\end{equation}
and so on.

In the particular case $\theta =\frac{1}{2}$, equation (\ref{fMasympt})
simplifies to 
\begin{equation}
f_{M}(x)=\frac{1}{\pi \sqrt{1-x^{2}}}=\beta (x),  \label{arcsin}
\end{equation}
which is the arcsine law on $[-1,1]$.

The corresponding distribution for the occupation time is 
\begin{equation}
\lim_{t\rightarrow \infty }f_{t^{-1}T^{\pm }}(x)=\frac{\sin \pi \theta }{\pi 
}\frac{x^{\theta -1}\left( 1-x\right) ^{\theta -1}}{x^{2\theta }+\left(
1-x\right) ^{2\theta }+2\cos \pi \theta \,x^{\theta }\left( 1-x\right)
^{\theta }},  \label{lamperti}
\end{equation}
a result originally found by Lamperti \cite{lamperti}. If $\theta =\frac{1}{2%
}$, the arcsine law on $[0,1]$ is recovered: 
\[
\lim_{t\rightarrow \infty }f_{t^{-1}T^{+}}(x)=\frac{1}{\pi \sqrt{x(1-x)}}.
\]

A last point is that, in the persistence region, 
i.e., for $M_{t}\rightarrow 1$ and $t\rightarrow\infty $,
$f_M(t;x)$ has a scaling form, as we now show.
Two limiting behaviors are already known:

\noindent (i) 
for $M_{t}=1$, $f_{M}(t;x)=\frac{1}{2}p_{0}(t)\delta (x-1)$,
with $p_{0}(t)\sim (t/\tau _{0})^{-\theta }$;

\noindent (ii) 
for $t=\infty $, the limiting distribution $f_{M}(x)$ is
given by equation (\ref{fM_x=1}).

\noindent
In order to interpolate between these two behaviors, we assume the scaling
form 
\[
f_{M}(t;x)\sim \left( \frac{t}{\tau _{0}}\right) ^{1-\theta }h\left( z=(1-x)%
\frac{t}{\tau _{0}}\right) , 
\]
in the persistence region defined above, with $z$ fixed. 
The known limiting behaviors imply 
\begin{equation}
h(z)\comport{\approx }{z\rightarrow \infty }2^{-\theta }\frac{\sin \pi
\theta }{\pi }\,z^{\theta -1},\qquad h(z)
\comport{\approx }{z\rightarrow 0}\frac{1}{2}\delta (z).  \label{scal_gz}
\end{equation}

The scaling function $h(z)$ can be computed in Laplace space as follows. We
have 
\[
\hat{f}_{S}(t;u)=\left\langle \hbox{e}^{-uS_{t}}\right\rangle =\left( \frac{%
\tau _{0}}{t}\right) ^{\theta }\hbox{e}^{-ut}\int_{0}^{\infty }dz\,\,\hbox{e}%
^{u\tau _{0}z}h(z). 
\]
Laplace transforming with respect to $t$, yields, in the limit $%
s+u\rightarrow 0$, with $u=O(1)$, since $z$ is finite, 
\[
\hat{f}_{S}(s;u)\approx a(s+u)^{\theta -1}\int_{0}^{\infty }dz\,\,\hbox{e}%
^{u\tau _{0}z}h(z). 
\]
By identification of this expression with the corresponding estimate
obtained in the same regime from (\ref{fS}), we have, with $v=-2u$, 
\[
\frac{1+\hat{\rho}(v)}{2(1-\hat{\rho}(v))}
=\int_{0}^{\infty }dz\,\,%
\hbox{e}^{-\frac{1}{2}v\tau _{0}z}h(z)=\hat{h}\left( \frac{v\tau _{0}}{2}%
\right) . 
\]
As a consequence, the function $h(z)$ is non universal since it depends on
the details of the function $\rho (\tau )$. Universality is restored only in
the two limits considered above (see (\ref{scal_gz})).

\medskip \noindent \textbf{(iii)} \textit{Broad distributions of intervals with index 
}$1<\theta <2$

Using equation (\ref{ro_broad}), (\ref{fS}) yields 
\begin{equation}
\hat{f}_{S}(s;u)\approx \frac{2\left\langle \tau \right\rangle -a\left(
(s+u)^{\theta -1}+(s-u)^{\theta -1}\right) }{2\left\langle \tau
\right\rangle s-a\left( (s+u)^{\theta }+(s-u)^{\theta }\right) },
\label{fSsu}
\end{equation}
in the scaling regime where both Laplace variables $s$ and $u$ are
simultaneously small and comparable. Expanding this expression as a Taylor
series in $u$, and performing the inverse Laplace transform term by term, we
obtain, for the first even moments of the random variable $S_{t}$, 
\begin{eqnarray}
&&\left\langle S_{t}^{2}\right\rangle 
\comport{\approx }{t\rightarrow \infty }\frac{2\tau _{0}^{\theta }}{(2-\theta )(3-\theta )\left\langle \tau
\right\rangle }t^{3-\theta },\qquad  \nonumber \\
&&\left\langle S_{t}^{4}\right\rangle 
\comport{\approx }{t\rightarrow \infty }\frac{4\tau _{0}^{\theta }}{(4-\theta )(5-\theta )\left\langle \tau
\right\rangle }t^{5-\theta }.  \label{mom_S}
\end{eqnarray}
This demonstrates that, in the long-time regime, the asymptotic distribution
of $S_{t}$ is broad, with slowly decaying tails. This distribution can be
evaluated as follows.

We first notice that equation~(\ref{fSsu}) further simplifies for 
$|u|\gg s$. In order for both variables of $\hat{f}_{S}(s;u)$ to stay in the
appropriate domains, we consider $s>0$ and $u=$i$\omega $, with $\omega $
real. We thus obtain, in the relevant regime $(s\ll |\omega |\ll 1)$, 
\[
\hat{f}_{S}(s;u)\approx \frac{1}{s+c|\omega |^{\theta }}, 
\]
with 
\[
c=-\frac{a}{\left\langle \tau \right\rangle }\cos (\pi \theta /2)=\frac{\pi
\,\tau _{0}^{\theta }}{2\Gamma (\theta )\sin (\pi \theta /2)\left\langle
\tau \right\rangle }. 
\]
In other words, we have the scaling $s\sim |u|^{\theta }$, or, for the
typical value of $S_{t}$, 
\[
(S_{t})_{\rm typ}\sim t^{1/\theta }. 
\]

The distribution of $S_{t}$ is then given, for $y$ small and $t$ large, by
the double inverse Laplace transform 
\[
f_{S}(t;y)\approx \int \frac{ds}{2\pi {\i}}\hbox{e}^{st}\int_{-\infty
}^{+\infty }\frac{d\omega }{2\pi }\frac{\hbox{e}^{{\i}\omega y}}{%
s+c|\omega |^{\theta }}\approx \int_{-\infty }^{+\infty }\frac{d\omega }{%
2\pi }\hbox{e}^{{\i}\omega y-c|\omega |^{\theta }t}. 
\]
Setting 
\begin{equation}
S_{t}=(ct)^{1/\theta }\,X_{t},  \label{def_X}
\end{equation}
we conclude that the scaling variable $X_{t}$ has a nontrivial even limiting
distribution, as $t\rightarrow \infty $, with argument 
$x=y/(c t)^{1/\theta}$, 
\begin{equation}
f_{X}(x)=L_{\theta }(x)=\int_{-\infty }^{+\infty }\frac{d\alpha }{2\pi }\,%
\hbox{e}^{{\i}\alpha x-|\alpha |^{\theta }}.  
\label{Lteta}
\end{equation}
This distribution is the symmetric stable L\'{e}vy law $L_{\theta }$ of
index $\theta $. Expanding the integrand in the right side as a Taylor
series in $x$ yields the convergent series 
\[
L_{\theta }(x)=\frac{1}{\pi \theta }\sum_{k=0}^{\infty }(-1)^{k}\,\frac{%
\Gamma \left( (2k+1)/\theta \right) }{(2k)!}\,x^{2k}. 
\]
For large values of $x$, $L_{\theta }(x)$ falls off as a power law, 
\[
L_{\theta }(x)\comport{\approx }{x\to \pm \infty }\frac{\Gamma (\theta
+1)\sin (\pi \theta /2)}{\pi }\,|x|^{-\theta -1}, 
\]
as obtained by expanding e$^{-|\alpha |^{\theta }}$ as $1-|\alpha |^{\theta
} $ in equation~(\ref{Lteta}). As a consequence, the second moment of this
distribution is divergent.

To summarize, the bulk distribution of $S_{t}$ is given by $L_{\theta }(x)$,
with the scaling (\ref{def_X}), while the moments of $S_{t}$ scale as (\ref
{mom_S}). 
These two kinds of behavior are actually related, as shown by the
following simple reasoning. 
Using (\ref{def_X}), we have 
\begin{equation}
\left\langle S_{t}^{2}\right\rangle \sim t^{2/\theta }\left\langle
X_{t}^{2}\right\rangle \sim t^{2/\theta
}\int_{-x_{c}}^{x_{c}}dx\,x^{2}\,L_{\theta }(x)\sim t^{2/\theta
}\,x_{c}^{2-\theta },
\end{equation}
where $x_{c}$ is an estimate for $x$ in the tails. Since $x_{c}\sim
S_{c}/t^{1/\theta }$ with $S_{c}\sim t$, by definition (\ref{def_S}) of $%
S_{t}$, we finally obtain $\left\langle S_{t}^{2}\right\rangle \sim
t^{3-\theta }$ as in (\ref{mom_S}). Note that, since $2/\theta <3-\theta $,
the typical value of $S_{t}^{2}$ is much smaller than its average, at large
times. 
A similar situation arises in a model considered in ref. \cite
{bouch-g}. 
We are indebted to J.P. Bouchaud for pointing this to us. 

This argument can be generalized to the calculation of non-integer
moments of $S_t$, yielding
$$
\left\langle |S_{t}|^{p}\right\rangle \sim t^{\gamma(p)},
$$
with $\gamma(p)=p/\theta$ if $p\le\theta$, and $\gamma(p)=p+1-\theta$
if $p\ge\theta$, a behavior characteristic of a  bifractal distribution.

In the limits $\theta \to 1$ and $\theta \to 2$, $L_{\theta}$ becomes
respectively a Cauchy and a Gaussian distribution, 
\[
L_{\theta }(x)\comport{\to }{\theta \to 1}
\frac{1}{\pi (1+x^{2})}%
,\qquad L_{\theta }(x)\comport{\to }{\theta \to 2}
\frac{\hbox{e}^{-x^{2}/4}}{2\sqrt{\pi }}. 
\]

\section{{\hskip -1.5pt}Number of renewals between two arbitrary times}

Consider the number of events $N(t,t+t')=N_{t+t'}-N_{t}$
occurring between $t$ and $t+t'$. The probability distribution of
this random variable is denoted by 
\[
p_{n}(t,t+t')=\mathcal{P}\left( N(t,t+t')=n\right) . 
\]

The time of occurrence of the $n$-th event, counted from time $t$, is
denoted by $t_{n}^{\prime }$, with, by convention, $t_{0}^{\prime }=0$. By
definition of the forward recurrence time $E_{t}$, the first event after
time $t$ occurs at time $t_{1}^{\prime }=E_{t}$, when counted from time $t$.
Hence the time of occurrence of the last event before $t+t'$,
counted from $t$, reads 
\[
t_{N(t,t+t')}^{\prime }=E_{t}+\tau _{2}+\cdots +\tau
_{N(t,t+t')}. 
\]
Therefore, 
\begin{equation}
p_{n}(t,t+t')=\mathcal{P}\left( t_{n}^{\prime }<t^{\prime
}<t_{n+1}^{\prime }\right) =\left\langle I(t_{n}^{\prime }<t^{\prime
}<t_{n+1}^{\prime })\right\rangle \qquad (n\geq 0).  \label{pntt'}
\end{equation}
In particular, for $n=0$, we have 
\begin{equation}
p_{0}(t,t+t')=\mathcal{P}(E_{t}>t')=\int_{t^{\prime
}}^{\infty }dy\,f_{E}(t;y),  
\label{p0tt'}
\end{equation}
which is the persistence probability up to time $t+t'$, counted
from the waiting time~$t$.

In Laplace space, where $u$ is conjugate to $t'$, we thus obtain 
\begin{eqnarray}
\lap{t'} p_{n}(t,t+t') 
&=&\hat{p}_{n}(t,u)=\hat{f}_{E}(t;u)\,\hat{\rho}(u)^{n-1}\frac{1-\hat{\rho}(u)}{u}%
\qquad (n\geq 1),  
\nonumber \\
\lap{t'}p_{0}(t,t+t') 
&=&\hat{p}_{0}(t,u)=\frac{1-\hat{f}_{E}(t;u)}{u},  \label{pntu}
\end{eqnarray}
where $\hat{f}_{E}(t;u)$ is the Laplace transform of $f_{E}(t;y)$ with
respect to $y$.

We discuss the results above according to the nature of the distribution of
intervals~$\rho(\tau)$.

\medskip \noindent \textbf{(i)} \textit{Narrow distributions of intervals}

As seen in section \ref{section_back}, the renewal process reaches
equilibrium at long times, with, for the forward recurrence time, 
\[
\hat{f}_{E,{\eq}}(u)
=\frac{1-\hat{\rho}(u)}{\left\langle \tau \right\rangle u},
\qquad f_{E,{\eq}}(y)
=\frac{1}{\left\langle \tau
\right\rangle }\int_{y}^{\infty }\,d\tau \,\rho (\tau )\,. 
\]
Therefore the equilibrium distribution of the random variable $%
N(t,t+t')$ no longer depends on $t$. In particular, the average of $%
N(t,t+t')$ is equal to $t'/\left\langle \tau \right\rangle 
$.

For an exponential distribution of time intervals, we have
\[
p_{n}(t,t+t')=
\hbox{e}^{-\lambda t'}\frac{(\lambda t')^{n}}{n!}
\qquad (n\geq 0), 
\]
which is independent of $t$, showing that the Poisson point process is at
equilibrium at all times.

\medskip \noindent \textbf{(ii)} 
\textit{Broad distributions of intervals with index }%
$\theta <1$

We restrict the discussion to the probability $p_{0}(t,t+t')$. 
In the scaling regime where $t$ and $t'$ are large and comparable, we
have, according to equations (\ref{p0tt'}) and~(\ref{fEscal}), 
\begin{eqnarray}
p_{0}(t,t+t') &\approx &\int_{t'/t}^{\infty
}dx\lim_{t\rightarrow \infty }\,f_{t^{-1}E}(x)=\frac{\sin \pi \theta }{\pi }%
\int_{t'/t}^{\infty }dx\frac{1}{x^{\theta }(1+x)}  
\nonumber \\
&=&\int_{0}^{t/(t+t')}dx\,\beta _{\theta ,1-\theta }(x)=g_{1}\left( 
\frac{t}{t+t'}\right) .
\label{p0tt'_l1}
\end{eqnarray}
In particular, in the regime of large separations between $t$ and 
$t+t'$, we obtain the aging form of the persistence probability: 
\[
p_{0}(t,t+t')\comport{\approx }{1\ll t\ll t'}
\frac{\sin \pi \theta }{\pi \theta }
\left( \frac{t'}{t}\right)^{-\theta }. 
\]
For example, for $\theta =\frac{1}{2}$, 
\begin{equation}
p_{0}(t,t+t')=\frac{2}{\pi }\arctan \sqrt{\frac{t}{t'}}=%
\frac{2}{\pi }\arcsin \sqrt{\frac{t}{t+t'}}.  \label{p0tt'scal}
\end{equation}

One can also compute 
\[
\lap{t,t'}\left\langle N(t,t+t^{\prime
})\right\rangle =\sum_{0}^{\infty }n\,\hat{p}_{n}(s,u)=\frac{\hat{\rho}(s)-%
\hat{\rho}(u)}{u(u-s)(1-\hat{\rho}(u))(1-\hat{\rho}(s))},
\]
or alternatively use (\ref{Nmean}), to find, in the same regime ($1\ll t\sim
t'$), 
\begin{equation}
\left\langle N(t,t+t')\right\rangle \approx \frac{\sin \pi \theta }{%
\pi \theta \,}\,\frac{(t+t')^{\theta }-t^{\theta }}{\tau
_{0}^{\theta }}.  \label{Ntt'l1}
\end{equation}
A consequence of (\ref{p0tt'_l1}) and (\ref{Ntt'l1}) is that, for $1\ll
t'\ll t$, the probability of finding an event between $t$ and $%
t+t'$ goes to zero. In other words, in order to have a chance to
observe a renewal, one has to wait a duration $t'$ of order $t$.
The intuitive explanation is that, as $t$ is growing, larger and larger
intervals of time $\tau $ may appear. The density of events at large times
is therefore decreasing.

\newpage
\noindent \textbf{(iii)} \textit{Broad distributions of intervals with index 
}$1<\theta <2$

First, equations (\ref{fBEeq}) and (\ref{p0tt'}) imply that, 
for $1\ll t'\ll t$, 
\begin{equation}
p_{0,{\eq}}(t,t+t')\approx 
\frac{\tau _{0}^{\theta }}{(\theta-1)\left\langle \tau \right\rangle }
t^{\prime -(\theta -1)}. 
\label{p0tt_eq}
\end{equation}
Then, in the scaling regime where $t$ and $t'$ are large and
comparable, we obtain, using equations (\ref{fE}) and (\ref{pntu}), 
\[
\hat{p}_{0}(s,u)\approx \frac{a}{\left\langle \tau \right\rangle }
\frac{s^{\theta -1}-u^{\theta -1}}{s(s-u)}, 
\]
which by inversion yields 
\begin{equation}
p_{0}(t,t+t')=\frac{\tau _{0}^{\theta }}{(\theta -1)\left\langle
\tau \right\rangle }\left( t^{\prime -(\theta -1)}-(t+t')^{-(\theta
-1)}\right) .  
\label{p0tt'_l2}
\end{equation}
For $1\ll t'\ll t$ we recover (\ref{p0tt_eq}), while for $1\ll t\ll t'$
we obtain
$$
p_{0}(t,t+t')\comport{\approx}{1\ll t\ll t'}
\frac{\tau _{0}^{\theta }t}{\left\langle \tau \right\rangle }
t'^{-\theta}.
$$

Equation (\ref{p0tt'_l2}) can be rewritten in scaling form as 
\begin{equation}
p_{0}(t,t+t')=\frac{\tau _{0}^{\theta }}{(\theta -1)\left\langle
\tau \right\rangle }t^{-(\theta -1)}g_{2}\left( \frac{t}{t+t'}%
\right) ,  \label{p0tt'_scal}
\end{equation}
with 
\begin{equation}
g_{2}(x)=x^{\theta -1}\left( \left( 1-x\right) ^{-(\theta -1)}-1\right) .
\label{gx}
\end{equation}

\section{Two-time autocorrelation function}
\label{auto}

The two-time autocorrelation function of the $\sigma $-process reads 
\[
C(t,t+t')=\left\langle \sigma_{t}\sigma_{t+t^{\prime
}}\right\rangle =\sum_{n=0}^{\infty }(-1)^{n}p_{n}(t,t'). 
\]
In Laplace space, using (\ref{pntu}), one gets 
\begin{eqnarray}
\lap{t'}C(t,t+t') &=&\hat{C}(t,u)=%
\frac{1-\hat{f}_{E}(t;u)}{u}-\hat{f}_{E}(t;u)\,\frac{1-\hat{\rho}(u)}{%
u\left( 1+\hat{\rho}(u)\right) }  \nonumber \\
&=&\frac{1}{u}\left( 1-\hat{f}_{E}(t;u)\frac{2}{1+\hat{\rho}(u)}\right) . 
\label{ctu1}
\end{eqnarray}

\newpage
\noindent \textbf{(i)} \textit{Narrow distributions of intervals}

At equilibrium, using (\ref{fBEeq1}), we have 
\[
\hat{C}_{{\eq}}(u)=\frac{1}{u}\left( 1-\frac{2\left( 1-\hat{\rho}%
(u)\right) }{\left\langle \tau \right\rangle u\left( 1+\hat{\rho}(u)\right) }%
\right) . 
\]
Expanding the right side as a Taylor series in $u$ yields the sum rules 
\begin{equation}
\int_{0}^{\infty }dt'C_{{\eq}}(t')=\frac{\left\langle
\tau ^{2}\right\rangle -\left\langle \tau \right\rangle ^{2}}{2\left\langle
\tau \right\rangle },\qquad \int_{0}^{\infty }dt't'\,C_{%
{\eq}}(t')=\frac{\left\langle \tau ^{3}\right\rangle }{%
6\left\langle \tau \right\rangle }-\frac{\left\langle \tau ^{2}\right\rangle 
}{2}+\frac{\left\langle \tau \right\rangle ^{2}}{4}.  \label{C_narrow}
\end{equation}
Since the second expression can be either positive or negative, depending on
the detailed form of the distribution $\rho (\tau )$, the equilibrium
correlation function $C_{{\eq}}(t')$ is neither positive nor
monotonic in general.

For an exponential distribution $\rho $, one has 
\[
C(t,t+t')=\hbox{e}^{-2\lambda t'} 
\]
which is independent of $t$, reflecting once again the fact that the process
is at equilibrium at all times.

\medskip \noindent \textbf{(ii)} \textit{Broad distributions of intervals}

In the regime where $s\sim u\ll 1$, the first term in
the first line of equation (\ref{ctu1}) 
dominates upon the second one for any $0<\theta <2$, so that we have 
\[
C(t,t+t')\approx p_{0}(t,t+t'), 
\]
with $p_{0}(t,t+t')$ given by (\ref{p0tt'_l1}) for $\theta <1$, and
by (\ref{p0tt'_l2}) or (\ref{p0tt'_scal}) for $1<\theta <2$.

Let us finally remark that, from the knowledge of the two-time
autocorrelation $C(t,t+t')$, we can recover the asymptotic
behavior of $\left\langle S_{t}^{2}\right\rangle $, or of $\left\langle
M_{t}^{2}\right\rangle $, respectively given by equations (\ref{S2_narrow}), (%
\ref{M2_broad1}) and (\ref{mom_S}), according to the nature of the
distribution of intervals $\rho $. 
Indeed, by definition of $S_{t}$ (see
equation (\ref{def_S})), we have 
\[
\left\langle S_{t}^{2}\right\rangle
=\int_{0}^{t}dt_{2}\int_{0}^{t}dt_{1}\,C(t_{1},t_{2}). 
\]
For a narrow distribution $\rho $, this yields, as $t\rightarrow \infty $, 
\[
\left\langle S_{t}^{2}\right\rangle 
\approx 2t\int_{0}^{\infty}dt'\,C_{\eq}(t'), 
\]
which, using (\ref{C_narrow}) leads to (\ref{S2_narrow}).
For a broad distribution $\rho $ with $\theta <1$, we have, in the long-time
scaling regime 
\begin{equation}
\left\langle S_{t}^{2}\right\rangle \approx 2t^{2}\int_{0}^{1}d\tau
_{2}\int_{0}^{\tau _{2}}d\tau _{1}\,C\left( \frac{\tau_{1}}{\tau _{2}}%
\right) =t^{2}\int_{0}^{1}dx\,g_{1}(x), 
\label{eq_var}
\end{equation}
where $g_{1}(x)$ is given in (\ref{p0tt'_l1}). 
A simple calculation then leads to $\langle M^2\rangle$ as in (\ref{M2_broad1}).
For a broad distribution $\rho $ with $1<\theta <2$, using (\ref{p0tt'_scal}%
), we find 
\[
\left\langle S_{t}^{2}\right\rangle \approx t^{3-\theta }\frac{2\tau
_{0}^{\theta }}{(\theta -1)(3-\theta )\left\langle \tau \right\rangle }%
\int_{0}^{1}dx\,x^{-(\theta -1)}\,g_{2}(x), 
\]
yielding, after integration, the first line of (\ref{mom_S}).

\section{Summary and final remarks}

This article is devoted to the study of the  occupation times 
$T_{t}^{\pm}$, and mean magnetization $M_{t}$, of renewal processes, that
is, processes with independent intervals of time between events, interpreted
as the zero crossings of the stochastic process $\sigma _{t}=\pm 1$. 
We also
compute the distributions of the random variables naturally associated to a
renewal process, such that the number of events $N_{t}$ occurring between $0$
and $t$, the epoch $t_{N}$ of the last event before $t$, the backward and
forward times $B_{t}$ and $E_{t}$. 
We finally investigate the number of
events occurring between two arbitrary instants of time, and the two-time
autocorrelation function.

The present work is also instructive for the understanding of the role of
correlations on the behavior of the distributions of sums of random
variables. Here the random variables are the signs 
$\sigma _{t}=\pm 1$, and 
\begin{equation}
S_{t}\equiv t\,M_{t}=\int_{0}^{t}dt^{\prime }\,\sigma _{t^{\prime }}
\end{equation}
is therefore the sum of temporally correlated random variables.

Three cases are to be considered for the discussion of the results,
according to the nature of the distribution $\rho (\tau )$ of the intervals
of time between the renewal events.

The case where $\rho $ is narrow corresponds to the domain of application of
the central limit theorem. Correlations between values of the sign process 
$\sigma_{t}$ at two different instants of time are short-ranged. All the
observables of interest mentioned above have narrow distributions, as well.
In particular, 
\[
(S_{t})_{\rm typ}\sim t^{1/2}, 
\]
and the limiting distribution of $S_{t}/t^{1/2}$ is Gaussian. The
persistence probability $p_{0}(t)$, i.e., the probability that no event
occurred up to time $t$, decreases, as $t\rightarrow \infty $, faster than a
power law.

The case where $\rho $ is broad, with index $\theta <1$, corresponds to a
maximum violation of the central limit theorem. The particular case where 
$\theta =\frac{1}{2}$ accounts for Brownian motion. Correlations are
long-ranged: 
\begin{equation}
\left\langle \sigma _{t}\sigma _{t+t^{\prime }}\right\rangle \sim
g_{1}\left( \frac{t}{t+t^{\prime }}\right) ,  
\label{conc_C1}
\end{equation}
where the scaling function behaves as $(t^{\prime }/t)^{-\theta }$ in the
regime of large separation ($1\ll t\ll t^{\prime }$). The number $N_{t}$ of
events between $0$ and $t$ scales as $t^{\theta }$. 
The random variables $t_{N}$, $B_{t}$, $E_{t}$, $T_{t}^{\pm }$ and 
$S_{t}$, have
limiting distributions, as $t\rightarrow \infty $, once rescaled by $t$. In
particular, focusing on $S_{t}$, the scaling behavior (\ref{conc_C1})
implies that 
\begin{equation}
(S_{t})_{\rm typ}\sim t.  \label{conc_Styp1}
\end{equation}
The limiting distribution of $S_{t}/t\equiv M_{t}$ is given by equation (\ref
{fMasympt}). It is a U-shaped curve for $\theta <\theta _{c}\approx 0.59$,
while its maximum is at zero, for $\theta >\theta _{c}$. It is singular at $%
M_{t}=\pm 1$, with an exponent equal to $\theta -1$. The probability, $%
p_{0}(t)$, that no event occurred up to time $t$, decreases as $t^{-\theta }$%
, while the probability, $p_{0}(t,t+t^{\prime })$, that no event occurred
between $t$ and $t+t^{\prime }$, behaves as $(t^{\prime }/t)^{-\theta }$ in
the regime $1\ll t\ll t^{\prime }$. (We remind that in the long-time scaling
regime, $\left\langle \sigma _{t}\sigma _{t+t^{\prime }}\right\rangle
\approx p_{0}(t,t+t^{\prime })$.)

The case where $\rho$ is broad, with index $1<\theta <2$, is intermediate.
Correlations are again long-ranged: 
\begin{equation}
\left\langle \sigma _{t}\sigma _{t+t^{\prime }}\right\rangle \sim
t^{-(\theta -1)}g_{2}\left( \frac{t}{t+t^{\prime }}\right) ,  \label{conc_C2}
\end{equation}
where the scaling function behaves as 
$(t'/t)^{-\theta}$ in the
regime of large separation ($1\ll t\ll t^{\prime }$). 
The law of $N_{t}$,
centered at $t/\left\langle \tau \right\rangle $, and rescaled by $%
t^{1/\theta }$, is broad, with index $\theta $. As $t\rightarrow \infty $, $%
B_{t}$ and $E_{t}$ have equilibrium distributions, which are broad, with
index $\theta -1$. The scaling behavior (\ref{conc_C2}) implies that 
\begin{equation}
\left\langle S_{t}^{2}\right\rangle \sim t^{3-\theta },  \label{conc_S2}
\end{equation}
while a more complete analysis leads to 
\begin{equation}
(S_{t})_{\rm typ}\sim t^{1/\theta }, 
\label{conc_Styp2}
\end{equation}
and shows that the limiting distribution of $S_{t}/t^{1/\theta }$ is the
symmetric stable L\'{e}vy law of index $\theta $. 
The existence of the two scales (\ref
{conc_S2}) and (\ref{conc_Styp2}) is also reflected in the bifractality of
the distribution of $S_{t}$, in the sense that  the non-integer moments
scale~as
\[
\left\langle |S_{t}|^{p}\right\rangle \sim t^{\gamma (p)}, 
\]
with $\gamma (p)=p/\theta $, if $p<\theta $, and $\gamma (p)=p+1-\theta $,
if $p>\theta $. 
Otherwise stated, there is no ``gap scaling'' for the
distribution of $S_{t}$ (see below). The behavior of $p_{0}(t)$ is the same
as for the previous case $\theta <1$. 
However, $p_{0}(t,t+t^{\prime })$,
though still decaying as $t'^{-\theta }$ in the regime $1\ll t\ll
t^{\prime }$, is no longer given by a scaling function of 
the ratio $t/t+t'$, as was the case when 
$\theta <1$ (compare equations (\ref{p0tt'_l1}) and (\ref{p0tt'_scal})).

The three types of behavior summarized above, corresponding to a narrow
distribution $\rho $ (i.e., formally $\theta >2$), to a broad distribution
with $\theta <1$, and to a broad distribution with $1<\theta <2$, are
reminiscent of the three typical behaviors observed in the nonequilibrium
dynamics of phase ordering, respectively at high temperature, at low
temperature, and at criticality. 
Consider for instance a two-dimensional
lattice of Ising spins $\sigma_{t}=\pm 1$. 
At high temperature,
correlations between spins are short-ranged, and the analysis given above
applies, i.e., the central limit theorem is obeyed by $S_{t}$. 
For a quench from a disordered initial state to a temperature
below the critical temperature, the autocorrelation has a 
form similar to (\ref{conc_C1}), namely
\[
\left\langle \sigma _{t}\sigma _{t+t^{\prime }}\right\rangle 
\sim m_{\eq}^2 \, g_{1}^{\rm Ising}\left( \frac{t}{t+t^{\prime }}\right),
\]
where $m_{\eq}$ is the equilibrium magnetization.
The scaling function $g_{1}^{\rm Ising}$ behaves as 
$(t'/t)^{-\lambda/z}$ for $1\ll t\ll t^{\prime }$, where
$z=2$ is the growth exponent, and $\lambda\approx 1.25$
is the autocorrelation exponent.
Correspondingly (\ref{conc_Styp1}) holds. 
The limiting distribution
of $S_{t}/t\equiv M_{t}$ is a U-shaped curve, with singularity exponent, as 
$M_{t}\rightarrow \pm m_{\eq}$, equal to $\theta -1$, where $\theta \approx 0.22$
is the persistence exponent at low temperature \cite{drouffe}.

Finally for a quench at the critical temperature, correlations have a form
similar to (\ref{conc_C2}),
\begin{equation}
\left\langle \sigma _{t}\sigma _{t+t^{\prime }}\right\rangle 
\sim t^{-2\beta/\nu z_c}
g_{2}^{\rm Ising}\left( \frac{t}{t+t^{\prime }}\right), 
\label{ctt_ising}
\end{equation}
where $\beta=1/8$ and $\nu=1$ are the usual static critical exponents,
and $z_c\approx 2.17$ is the dynamic critical exponent.
The scaling function $g_{2}^{\rm Ising}$ behaves as 
$(t'/t)^{-\lambda_c/z_c}$ for $1\ll t\ll t^{\prime }$, where
$\lambda_c\approx 1.59$
is the critical autocorrelation exponent (see e.g. ref. \cite{g-l}). 
However, in contrast with the
situation encountered in the present work, one expects, by analogy with statics,
that the distribution
of $S_{t}$ is entirely described by one scale, given by 
$(S_{t})_{\rm typ}=\left\langle S_{t}^{2}\right\rangle^\frac{1}{2}
\sim t^{1-\beta/\nu z_c}$. 
In this sense, this distribution is a
monofractal, and gap scaling holds: all moments scale as 
$\left\langle |S_{t}|^{p}\right\rangle 
\sim (S_{t})_{\rm typ}^p
\sim t^{\gamma(p)}$, 
with $\gamma(p)=p(1-\beta/\nu z_c)$.
These expectations are confirmed by numerical computations~\cite{drouffe3}.

Pursuing the comparison between the two situations,
persistence for coarsening systems decays faster than a power law
at high temperature, and as a power law at low temperature, as is the case
for a renewal process, respectively for $\rho$ narrow, and for $\rho$ broad,
with $\theta<1$. 
Again the status of the intermediate case is
different: while for the renewal processes under study 
$p_{0}(t)$ is still decaying as 
$t^{-\theta }$ when $1<\theta <2$, for critical coarsening the decay of the
probability  that no spin flip occurred up to time $t$, $p_{0}(t)$, is no
longer algebraic.

To conclude, as we already mentioned, the methods introduced here provide a
basis for the study of similar questions for the stochastic process
considered in ref.~\cite{dhar-majumdar}, and to some extent, for the random
acceleration problem \cite{ustocome2}.

\bigskip

After completion of this article,
we became aware of recent works in probability theory
on various quantities derived from Brownian motion,
whose distributions are proven to be beta laws,
generalizing thus L\'evy's arcsine law~\cite{yor}.

\subsection*{Acknowledgments}

We thank an anonymous referee for pointing out the connection
of the present article with recent works in probability theory,
and M. Yor and F. Petit for introducing us to these works.

\newpage

\appendix

\section{A word on notations}

\noindent \textit{Probability densities}

Consider the time-dependent random variable $Y_{t}$. We are interested in
the distribution function of this random variable, $\mathcal{P}(Y_{t}<y)$,
and in its probability density function, denoted by $f_{Y}$, dropping the
time dependence of the random variable when it is in subscript, 
\[
f_{Y}(t;y)=\frac{d}{dy}\mathcal{P}(Y_{t}<y).
\]
Time $t$ appears as a parameter in this function.\medskip 

\noindent \textit{Laplace transforms}

Assume that $Y_{t}$ is positive. We denote the Laplace transform of $%
f_{Y}(t;y)$ with respect to $y$ as 
\[
\lap{y}f_{Y}(t;y)=\hat{f}_{Y}(t;u)=\left\langle \hbox{e}%
^{-uY_{t}}\right\rangle =\int_{0}^{\infty }dy\,\hbox{e}^{-uy}\,f_{Y}(t;y), 
\]
and the double Laplace transform of $f_{Y}(t;y)$ with respect to $t$ and $y$
as 
\[
\lap{t,y}f_{Y}(t;y)=\lap{t}%
\left\langle \hbox{e}^{-uY_{t}}\right\rangle =\hat{f}_{Y}(s;u)=\int_{0}^{%
\infty }dt\,\hbox{e}^{-st}\int_{0}^{\infty }dy\,\hbox{e}^{-uy}\,f_{Y}(t;y). 
\]

In this work we encounter random variables $Y_{t}$ (such as $S_{t}$ or $%
M_{t} $) with even distributions on the real axis, i.e., $f_{Y}(t;y)=$ $%
f_{Y}(t;-y) $. 
For these we define the (bilateral) Laplace transform as 
\[
\lap{y}f_{Y}(t;y)=\left\langle \hbox{e}%
^{-uY_{t}}\right\rangle =\int_{-\infty }^{\infty }dy\,\hbox{e}%
^{-uy}\,f_{Y}(t;y). 
\]
\medskip

\noindent \textit{Limiting distributions}

In a number of instances considered in this work, $Y_{t}$ scales 
asymptotically as~$t$.
Therefore, it is natural to define the scaling variable $X_{t}=Y_{t}/t$,
with density $f_{X}(t;x)\equiv f_{t^{-1}Y}(t;x=y/t)$. 
As $t\rightarrow \infty $, this density converges to a limit, denoted by 
\[
f_{X}(x)= \lim_{t\rightarrow \infty }f_{t^{-1}Y}(x). 
\]

These considerations hold similarly for other scaling forms of $Y_{t}$.

\section{Inversion of the scaling form of a double Laplace transform}

Consider the probability density function $f_{Y}(t;y)$ of the random
variable $Y_{t}$, and assume that its double Laplace transform with respect
to $t$ and $y$, defined in appendix~A, has the scaling behavior 
\begin{equation}
\lap{t,y}f_{Y}(t;y)=\hat{f}_{Y}(s;u)=\frac{1}{s}\,
g\!\left(\frac{u}{s}\right)  
\label{hypothese}
\end{equation}
in the regime $s,u\rightarrow 0$, with $u/s$ arbitrary. Then the following
properties hold, as shown below.

\noindent (i) The random variable $X_{t}=Y_{t}/t$ possesses a limiting
distribution when $t\rightarrow \infty $, i.e., 
\begin{equation}
f_{X}(x)=\lim_{t\rightarrow \infty }f_{t^{-1}Y}(t;x=y/t).  \label{a}
\end{equation}

\noindent (ii) The scaling function $g$ is related to $f_{X}$ by 
\begin{equation}
g(\xi )=\left\langle \frac{1}{1+\xi X}\right\rangle =\int_{-\infty }^{\infty
}dx\,\frac{f_{X}(x)}{1+\xi x}.  \label{b}
\end{equation}

\noindent (iii) This can be inverted as 
\begin{equation}
f_{X}(x)=-\frac{1}{\pi x}\lim_{\epsilon \rightarrow 0}\Im g\left( -%
\frac{1}{x+{\i}\epsilon }\right) .  \label{c}
\end{equation}
(iv) Finally the moments of $f_{X}$ can be obtained by expanding $g(y)$ as a
Taylor series, since (\ref{b}) implies that
\begin{equation}
g(\xi )=\sum_{k=0}^{\infty }(-\xi )^{k}\left\langle X^{k}\right\rangle .
\label{d}
\end{equation}

First, a direct consequence of the scaling form (\ref{hypothese}) is that 
$Y_{t}$ scales as $t$, as can be seen by Taylor expanding the right side of
this equation, which generates the moments of $Y_{t}$ in the Laplace space
conjugate to $t$. 
Therefore (\ref{a}) holds.

Then, (\ref{b}) is a simple consequence of (\ref{a}), since 
\[
\hat{f}_{Y}(s;u)=\int_{0}^{\infty }dt\,\hbox{ e}^{-st}\left\langle \hbox{e}%
^{-uY}\right\rangle =\int_{0}^{\infty }dt\,\hbox{e}^{-st}\left\langle \hbox{e%
}^{-ut\,X}\right\rangle =\left\langle \frac{1}{s+uX}\right\rangle . 
\]
Now, 
\[
f_{X}(x)=\left\langle \delta \left( x-X\right) \right\rangle =-\frac{1}{\pi }%
\lim_{\epsilon \rightarrow 0}\Im\left\langle \frac{1}{x+{\i}%
\epsilon -X}\right\rangle . 
\]
The right side can be rewritten using (\ref{b}), yielding (\ref{c}).

\end{document}